# A Study of Influential Factors in the Adoption and Diffusion of B2C E-Commerce

Rayed AlGhamdi (ال غامدي رائد د)
Faculty of Computing and IT
King Abdulaziz University
Jeddah, Saudi Arabia

Ann Nguyen
School of ICT, Griffith University
170 Kessels Rd, Nathan
QLD 4111, Australia

Vicki Jones
School of ICT, Griffith University
170 Kessels Rd, Nathan
QLD 4111, Australia

***Abstract*—This paper looks at the present standing of e-commerce in Saudi Arabia, as well as the challenges and strengths of Business to Customers (B2C) electronic commerce. Many studies have been conducted around the world in order to gain a better understanding of the demands, needs and effectiveness of online commerce. A study was undertaken to review the literature identifying the factors influencing the adoption and diffusion of B2C e-commerce. It found four distinct categories: businesses, customers, environmental and governmental support, which must all be considered when creating an e-commerce infrastructure. A concept matrix was used to provide a comparison of important factors in different parts of the world. The study found that e-commerce in Saudi Arabia was lacking in Governmental support as well as relevant involvement by both customers and retailers.**

## I. Introduction

Saudi Arabia has a large and growing ICT marketplace. Yet, despite its size and rapid growth, progress in e-commerce activities is relatively slow [1-3]. The Saudi Government introduced e-commerce in 2001 in response to the fast expansion of e-commerce throughout the world. A permanent technical committee for e-commerce was established by the Saudi Ministry of Commerce. However, this Committee no longer exists, and from 2006, e-commerce supervision and development has been managed by the Ministry of Communications and Information Technology (MCIT). Unfortunately, there has been little progress since then [4].

## II. Factors Influencing the Adoption and Diffusion of B2C E-Commerce

Around the world, many studies in online commerce have been conducted in order to gain a better understanding of its strengths and challenges. Research into the factors influencing the adoption and diffusion of B2C e-commerce tends to discuss these factors as belonging to one or more of four categories. These include: businesses, customers, environment and government facilitation.

### A. *Factors influencing business' adoption of B2C e-commerce*

The literature discusses various issues that influence businesses to adopt B2C e-commerce. The highlighted issues in this review include

- organization e-readiness
- competitive pressure
- set-up and maintenance cost
- brand strength
- relative advantage of using e-commerce
- consumer purchasing power
- Privacy and Security
- type of products
- Resistance to change

### B. *Factors influencing customers to purchase online*

The literature discusses various issues that influence consumers to purchase online. The highlighted issues in this review include

- lack of trust due to security/privacy concerns
- reluctance to use credit cards
- language barriers
- preferences for in-store shopping
- good quality of e-commerce websites
- lack of product trial / inspect by hand
- relative advantage (prices, convince etc)
- familiarity of products/seller's good reputation

### C. *Environmental factors influencing the rate of B2C e-commerce adoption and diffusion*

Environmental factors affect the online environment and e-commerce activities, and as a result, also affect businesses and customers. This means that these factors are influential when businesses choose to adopt e-commerce and when customers decide to start trading using e-commerce. The most highlighted issues include

- ICT infrastructure
- online payment mechanisms
- the degree of credit cards penetration
- legislative and regulatory framework
- logistics Infrastructure
- education and awareness

### D. *Government intervention and its influence to the rate of B2C e-commerce adoption*

Throughout the World, governments tend to encourage and support the development of e-commerce. Also, as governments gain a larger Internet presents, their role often changes from simply informational to transactional. As a result, the government becomes both supplier and consumer, therefore contributing to the growth of e-commerce [5]. Using e-commerce also means that government departments can reach more customers, with faster service in a more economical way [6].

In Saudi Arabia, there are relatively few studies which identify business, consumers, and government factors. Most of the studies conducted, concentrate on environmental factors. Similarly with Developing Nations and the Gulf Region, the majority of studies have also focused on the environmental factors. However, studies conducted in Developed Countries tend to concentrate mainly on business and, to a lesser degree, government and environmental factors, with a smaller percentage on customers. This correlates with the fact that the environment, or e-commerce infrastructure, is already well established in these countries.

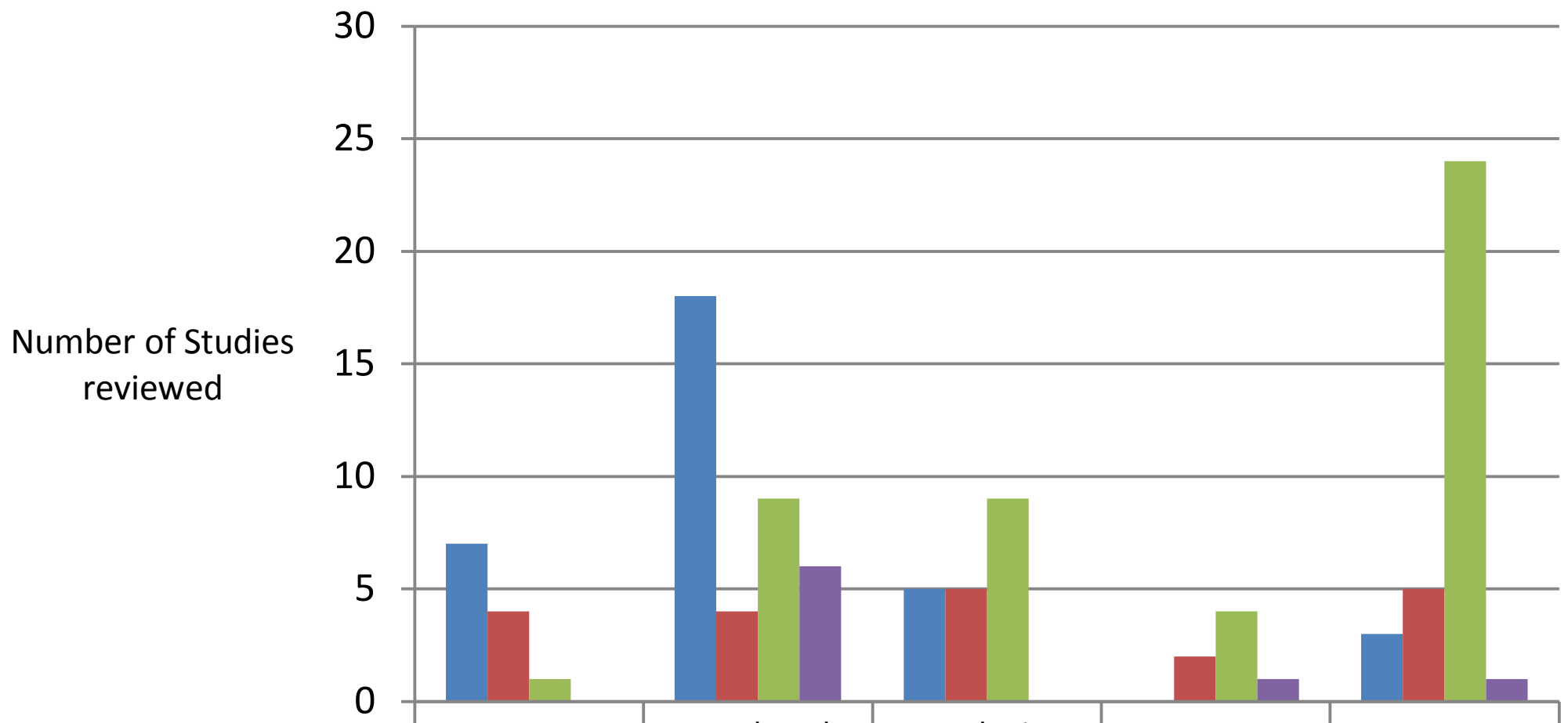

| | Not specific | Developed Countries | Developing Countries | Gulf Region | Saudi Arabia |
|---|---|---|---|---|---|
| Business | 7 | 18 | 5 | 0 | 3 |
| Customers | 4 | 4 | 5 | 2 | 5 |
| Environment | 1 | 9 | 9 | 4 | 24 |
| Government | 0 | 6 | 0 | 1 | 1 |

Figure 1. Factors and associated Studies which have influenced the adoption and diffusion of B2C e-commerce

The factors, which have influenced the adoption and diffusion of B2C e-commerce in research literature, are summarized in Table I (see the appendix). These factors and associated studies are further identified in Figure 1.

A study was undertaken to review the literature identifying the factors influencing the adoption and diffusion of B2C e-commerce. The review is divided into four sections: businesses, customers, environmental and governmental factors (see Table I, in the appendix).

A concept-centric structure is used to enable the separate influencing factors to be associated with the geographical context of the studies reviewed. The resulting concept matrix was then be used to provide a comparison of important factors in different parts of the world. The studies covered the geographical locations of: not specific, developed countries, developing countries, Gulf countries and Saudi Arabia. This division helps to compare different nations and identify similarities and differences (Figure 1).

In regards to Developing Nations, the Gulf Region and Saudi Arabia, the environment (e-commerce infrastructure) is not yet fully established, so is an important factor influencing the adoption and diffusion of B2C e-commerce. These figures also suggest that governmental support should be a high priority for e-commerce development. By contrast, the high level of readiness in e-commerce environment in Developed Countries leads most studies to concentrate on the businesses and why they might not be selling online when the e-commerce environment is ready for them.

## III. E-GOVERNMENT IN SAUDI ARABIA

E-government and e-commerce share some similarity in terms of transaction requirements. Therefore, development in e-government can serve as an engine to power e-commerce development [5]. The similarity between e-government and e-commerce is that both of them depend on ICT infrastructure, online payment systems and mailing/post systems to reach their users/customers and deliver their services/products [7,8].

In 2003, a decision was taken by the Saudi Government to start work on e-government; however, a committee for e-government was established and the actual work started in 2005 [9,10]. With cooperation of three government entities (The Ministry of Communication and Information Technology (MCIT), the Ministry of Finance and the Communication and IT Committee (CITC)), an e-government program called 'Yasser' was launched in 2005 [7]. This program acts as an umbrella for all e-government activities, procedures, legislations and all related issues [10]. An e-government plan was set up with the following vision "By the end of 2010, everyone in the Kingdom will be able to enjoy from anywhere and at any time – world class government services offered in a seamless user friendly and secure way by utilizing a variety of electronic means" [7]. However, this vision has not been achieved as set up in a timely manner, which means the plan was not realistic [10]. The main problem, which was not taken into consideration, was the ICT infrastructure and assessing the e-readiness of the different government departments [11-14]. As a result, an e-government second action plan with the vision: "Enable use of efficient, integrated customer friendly and secure multiple e-Government services" (covering the period 2012-2016) has been launched; considering human resource training and development, promote cooperation and innovation culture, and maximizing efficiency of e-services provided by government agencies,[15].

## IV. Government role in e-commerce promotion

Government support takes various forms from country to country; however, government regulation can be critical to supporting e-commerce [16]. Online shopping shows rapid growth in the developed world. Significantly, the South Korean government has played a key role promoting e-commerce. The Malaysian government is encouraging small and medium enterprises (SMEs) to adopt e-commerce solutions, and in Australia, the Government is providing support is various forms [17].

Although Australian online shoppers are increasing, many prominent Australian retailers are still lagging in terms of online sales. An Australian Government report about e-commerce stated that many retailers did not understand the potential benefits of online shopping and were concerned about the set-up and maintenance costs. The government followed by hosting an online retail forum to encourage, assist and inform retailers [17]. Small Business Online (SBO) was announced in the 2009-10 budget granting $14 million to help small businesses go online by offering training programs, advice and development of e-business resources [18].

The Government program, AusIndustry, supports business programs and a range of incentives aimed at helping businesses grow [18]. Singapore, South Korea and Hong Kong are also good examples of countries where the government has taken an active role in pushing for e-commerce proliferation [19]. The Singaporean government has been supporting e-commerce in the country since the early '90s. South Korea has a very strong ICT infrastructure, and Hong Kong set up and implemented an Electronic Transactions Ordinance in 2000. The Hong Kong ordinance contributed to future growth in e-commerce by providing for a legal infrastructure, such as the use digital signatures [20].

## V. Discussion/Conclusion

From the study results, it seems that Saudi Arabia is in great need of more Governmental support. The other two areas which need to be increased are the customers and the retailers. Those countries with successful online retailing infrastructures have strong Governmental support. In Saudi Arabia, people tend to feel more confident in business ventures if they are backed by the Government [21]. It would seem logical that once the Government is involved, the element of trust that ensues can encourage customers and retailers to become involved.

A strategic plan needs to be developed to promote online retailing in Saudi Arabia. Based on this research it is apparent that e-commerce in Saudi Arabia is still in its early stages. With Government engagement, the current state of online retailing could transform into a fully integrated online retailing infrastructure. An appropriate method will be proposed in a later publication.

TABLE I. FACTORS WHICH HAVE INFLUENCED THE ADOPTION AND DIFFUSION OF B2C E-COMMERCE

| Studies / Factors | Not specific | | | | Developed Countries | | | | | | | | | Developing countries | | | | | | | Gulf Region | | Saudi Arabia | | | | | | | | |
|---|---|---|---|---|---|---|---|---|---|---|---|---|---|---|---|---|---|---|---|---|---|---|---|---|---|---|---|---|---|---|---|
| | [22] | [23] | [24] | [25] | [26] | [27] | [27] | [28] | [29] | [30] | [31] | [32] | [33] | [34] | [35] | [36] | [37] | [38] | [39] | [19] | [40] | [41] | [42] | [43] | [44] | [1] | [45] | [46] | [47] | [48] | [49] |
| **Business** | | | | | | | | | | | | | | | | | | | | | | | | | | | | | | | |
| Organization e-readiness | | ● | ● | | | | ● | | ● | ● | | | | ● | ● | | | | | | | | | | | | | | | ● | |
| Competitive pressure | | | ● | | | | ● | | | ● | | | | ● | | | | | | | | | | | | | | | | | |
| Set-up and maintenance cost | | | | | | | | | | ● | | | | | | | | | | | | | | | | | | | | | |
| Brand strength | | | | ● | | | | | | | | | | | | | | | | | | | | | | | | | | | |
| Relative advantage of using e-commerce | ● | | ● | | ● | | ● | ● | | ● | ● | | | ● | | | | | | | | | | | | | | | ● | | |
| Consumer purchasing power | | | | | | | ● | ● | | | ● | | | | ● | | | | | | | | | | | | | | ● | | |
| Privacy and Security | | | | | | | | | | ● | ● | | | | | | | | | | | | | | | | | | | | |
| Type of products | | ● | | | | | | | | | | | | | | | | | | | | | | | | | | | | | |
| Resistance to change | | | | | | | ● | ● | | | | | | | | | | | | | | | | | | | | | | | |
| **Customers** | | | | | | | | | | | | | | | | | | | | | | | | | | | | | | | |
| Lack of trust due to security/privacy concerns | | ● | | | | | | ● | | | | | | | ● | | | | | ● | | ● | | | | | | ● | | | ● |
| Reluctance to use credit cards | | | | | | | | | | | | | | | | | ● | | | | | | | | | | ● | ● | | | |
| Language barriers | | | | | | | | ● | | | | | | | | | | | | | | | | | | | | | | | |
| Preferences for in-store shopping | | | | | | | | ● | | | | | | | | | | | | | | | | | | | | | | | |
| Good quality of e-commerce websites | | ● | | | | | | ● | | | | | | | | | | | | | | ● | | | | | | | | | ● |
| Lack of product trial / inspect by hand | | ● | | | | | | | | | | | | | | | | | | | | | | | | | | | | | |
| Relative advantage (prices, convince etc) | | ● | | | | | | | | | | | | | | | | | | | | | | | | | | | | | |
| Familiarity | | | | | | | | | | | | | | | ● | | | | ● | | | | | | | | | | | | |
| of products/seller's + reputation | | | | | | | | | | | | | | | | | | | | | | | | | | | | | | | |
| **Environment** | | | | | | | | | | | | | | | | | | | | | | | | | | | | | | | |
| ICT infrastructure | | | | | | | | ● | | | | ● | | ● | ● | ● | | ● | | ● | | ● | ● | ● | | ● | | ● | ● | ● | |
| Online payment mechanisms | | | | | | | | ● | | | | | | | | ● | ● | | | | | ● | | | | ● | ● | | | ● | |
| The degree of credit cards penetration | | | | | | | | | | | ● | ● | | | | | | | | | | | | | | | ● | | | | |
| Legislative and regulatory framework | | | | | | | | | ● | | ● | | | | | ● | | | | | | ● | ● | | ● | | | ● | ● | ● | |
| Logistics Infrastructure | | ● | | | | | | | | | | | | | | | | | | | | | | ● | ● | ● | ● | ● | | ● | |
| Education and Awareness | | | | | | | | | ● | | | ● | | | | | | | | ● | | ● | | ● | | | ● | | ● | | |
| **Government** | | | | | | | | | | | | | | | | | | | | | | | | | | | | | | | |
| Government intervention/ promotion | | | | | | ● | | ● | ● | ● | ● | | ● | | | | | | | | ● | | | | | | | | ● | | |